# Structural and phase properties of tetracosane ($C_{24}H_{50}$) monolayers adsorbed on graphite: an explicit hydrogen Molecular Dynamics study

L. Firlej[1,3], B. Kuchta[2,3], M.W. Roth[3], M.J. Connolly[1], and Carlos Wexler[4]

[1]*LCVN, Université Montpellier 2, 34095 Montpellier, France*
[2]*Laboratoire Chimie Provence, Université de Provence, 13396 Marseille, France*
[3]*Department of Physics, University of Northern Iowa, Cedar Falls, IA 50614, USA*
[4]*Department of Physics and Astronomy, University of Missouri, Columbia, MO 65211, USA*

**Abstract**

We discuss Molecular Dynamics (MD) computer simulations of a tetracosane ($C_{24}H_{50}$) monolayer physisorbed onto the basal plane of graphite. The adlayer molecules are simulated with explicit hydrogens, and the graphite substrate is represented as an all-atom structure having six graphene layers. The tetracosane dynamics modeled in the fully atomistic manner agree well with experiment. The low-temperature ordered solid organizes in rectangular centered structure, incommensurate with underlying graphite. Above $T$ = 200 K, as the molecules start to lose their translational and orientational order via gauche defect formation, a weak smectic mesophase (observed experimentally but never reproduced in United Atom (UA) simulations) appears. The phase behavior of the adsorbed layer is critically sensitive to the way the electrostatic interactions are included in the model. If the electrostatic charges are set to zero (as it is in UA force field), the melting temperature increases by ~70 K with respect to the experimental value. When the non-bonded 1-4 interaction is not scaled, the melting temperature decreases by ~90 K. If the scaling factor is set to 0.5, the melting occurs at $T$ = 350 K, in very good agreement with experimental data.

**Introduction**

Computer studies of thermodynamic properties of intermediate-length alkanes physisorbed on solid substrates have for many reasons become a very active research field. From a practical point of view, interfacial properties of n-alkanes are relevant to many wide-spread applications and problems (e.g. lubrication, adhesion, catalysis, etc.). Alkanes can also be considered as simplified model materials which help to understand similar but larger systems such as polymers, lipids or proteins. In fact, on the one hand they possesses a degree of complexity and are large enough to exhibit features relevant to linear molecules and, on the other hand, they are still sufficiently small to be tractable in computer simulations such as Molecular Dynamics or Monte Carlo methods.

Nearly all extensive computer simulations of adsorbed alkane layers employ the United Atom (UA) approximation. In this approach, methyl ($CH_3$) and methylene ($CH_2$) are represented as spherical pseudoatoms that replace the respective groups in the molecule. Of course, such an approximation saves a considerable amount of computational time and has led in the past to valuable progress in our understanding of structure and dynamics of the adsorbed alkane layers (at microscopic level). In particular, UA-based MD simulations have provided important informa-tion on the mechanism of melting of intermediate-length n-alkanes on graphite. They have shown evidence that the rotation of molecules out of adsorption plane during the monolayer melting (the "footprint" reduction) leads to a simultaneous loss of both intramolecular and translational order in the system. Such a cooperative process is related to an abrupt increase in number of *gauche* defects within the central region of the alkane chain[1]. Such an observation seems to be a key feature of melting. It suggests that melting occurs because of a strong correlation between deformations of the alkane (so called 'chain melting') and the translational disorder (lattice melting)[2] As a consequence, the melting temperature must strongly depend on the molecular stiffness,[1,2] which also plays a central role in this work and is discussed later in this paper.

The UA approximation has, however, several important shortfalls. Some of them directly result from the neglecting of the exact hydrogen location along the molecular chain, and they have been enumerated in the recent paper by Connolly *et al.*[3] In particular, the authors have shown that the lack of explicit hydrogen atoms leads to an unrealistic rolling of molecules out of adsorption plane. Such a behavior may result in an erroneous analysis of both structural and phase behavior of the system.[3]

In addition, in many situations when UA model has been used, electrostatic terms in energy calculations have been totally neglected[4,5]. Such a simplification seems to be justified according to the recent paper of Song *et al.*[6] that showed that electrostatic interactions have a negligible impact on most properties of alkanes. However, *ab initio* calculations indicate significant charge separation in C-H bonds (0.3-0.6 D). According to that result, the OPLS-AA models include partial charges on both carbon and hydrogen in the molecules. Although these non-negligible bond dipoles noticeably affect the interaction energies for a given configuration of two molecules, orientational averaging tends to reduce their importance. In larger alkanes, when symmetry is lower than tetrahedral, one might expect electrostatics to have a more significant influence.[6]

There are some reports showing that electrostatic and polarization energies appear to be of similar importance.[7] In fact, even if effective pairwise electrostatic potentials between static charges are widely used and have been successful in simulations of condensed-phase systems, there is a continuing effort to incorporate many-body polarization effects to further improve the accuracy of the force fields.[8] In this work, we use the standard CHARMM charges distributed over carbon and hydrogen atoms in methyl and methylene groups. The polarization energy is not taken into account.

In this paper we analyze the dynamic behavior of a tetracosane ($C_{24}H_{50}$) monolayer adsorbed on graphite. First of all, using an all-atom model of molecules, we shed new light on the structural phase transitions occurring within the adsorbed film. In particular, melting is preceded by reorganization of layer into a smectic-like structure. It is the first time that numerical results are able to reproduce corresponding experimental (neutron) measurements by Taub *et al*[9] - something never accomplished with UA simulations. In addition, we show that an appropriate modeling of molecular stiffness through scaling of internal non-bonded energy terms is essential for reproducing experimental melting temperatures. Specifically, the electrostatic force (restricted to interaction between point charges on the atoms) plays an important role in defining the melting mechanism. The results have already motivated further study of the influence of the polarization term on structure and dynamics of intermediate-length alkane layers effect which will be reported in a future paper [12].

**Simulation model**

The (*NVT*) canonical ensemble molecular dynamics (MD) method is used to simulate the tetracosane / graphite system. The simulation box of dimensions ($X = 34 \times 4.26$ Å, $Y = 53 \times 2.46$ Å) is commensurate with graphite lattice. Periodic boundary conditions are implemented in the ($x,y$) plane and free boundary conditions are applied in the vertical ($z$) direction. The graphite substrate is modeled in an all-atom fashion, as six identical sheets stacked in A-B-A-B graphite pattern. The tetracosane molecular structure and charge distribution over the atoms are taken from the Brookhaven Protein Data Bank (PDB). All carbons and hydrogen have been explicitly included in the calculations. The number of molecules in the simulation box is $N = 128$. In the initial (low temperature) configuration, the molecules form parallel lamellae arranged in rectangular centered structure. The initial configuration for the system is shown in Figure 1.

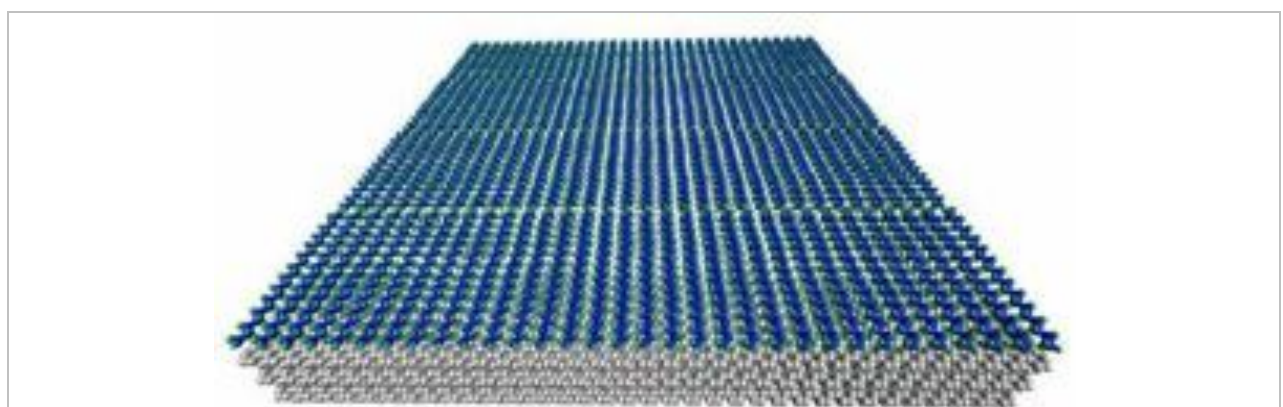

**Figure 1**. Initial configuration for the C24 / gr system. Adsorbate carbons are blue, hydrogens are green and graphite carbons are gray.

The interaction model applied in this paper uses the CHARMM22 format and parameters. Non-bonded interactions include an electrostatic interaction, a two-body atom-atom Lennard-Jones interaction between different molecules, and atom-substrate pair interactions. The internal degrees of freedom include the three-body bond angle bending and four-body dihedral torsions. The values of potential parameters are listed in Ref. 3 (Tables 2 and 3). The MD simulations were performed using parallel NAMD2 program[10] running on a Linux cluster. The time step in the simulations was 1 fs. The thermal averages are calculated over 3-5 ns time interval in equilibrated runs.

Correctly accounting for the intra-molecular terms in the interaction model is extremely important *vis–a–vis* the interplay between the bonded and non-bonded interactions. In fact, intra-molecular interaction terms directly define the overall stiffness of the molecules, which has a crucial effect on the mechanism of melting[1,2] We have used the scaled 1-4 exclusion policy implemented in the CHARMM force field. It assumes that all 1-2 and 1-3 pairs of atoms (that is, atoms separated by one or two bonds in a 1-2-3-4-5-…chain) are excluded from calculations of non-bonded interactions. Only the internal, bond energy is included. For 1-5 and longer distances, both electrostatic and van der Waals (vdW) energies are calculated explicitly. At intermediate distances, for atoms separated by three bonds (1-4 interactions) the explicit non-bonded part is scaled down as it is already partially included in dihedral angle torsion term. However, the value of the scaling factor is subject to debate. For now, the scaling factor has a physical justification but no physically established values. This important issue will be discussed elsewhere [12].

Here, we consider three limiting situations. Starting from the non-scaled model (1-4 scaling factor SF = 1, van der Waals parameters not modified), we first totally suppressed the electrostatic interaction (as in the UA approximation). As will be shown below, such modifications of the interactions shift the melting temperature of tetracosane monolayer adsorbed on graphite approximately by ~150 K, showing how sensitive is the dynamic behavior to the electrostatic interaction. Therefore, in the final analysis we use the scaled 1-4 non-bonded electrostatic contribution (using SF = 0.5, which is a typical value suggested for simulations of long molecules, polymers and proteins) and the van der Waals interactions are represented by the special 1-4 parameters as defined in CHARMM22 force field.

**Results**

**1. Melting transition.**

The melting of adsorbed layers of alkanes is a process that is more complicated than in conventional fluids. The main factor responsible for the mechanism is the intricate competition between intermolecular and intramolecular forces in the presence of substrate. Therefore, it is crucial to have an interaction model where the balance between those two components is correct; otherwise, the simulated picture may be even qualitatively wrong. The CHARMM force field applied in this paper is known for its realistic interaction parameters, but also for having some drawbacks which must be tested in specific situations. The results presented in this paper show how influential the balance between different terms of interaction model are if one wants to understand the mechanism of phase changes in the tetracosane layers whose molecules are not rigid.

The most delicate point in the CHARMM force field is the parametric adjustment of the 1-4 dihedral bond distortion terms. Indeed, the 1-4 non-bonded interaction influences the overall stiffness of molecules. The precise value of dihedral distortion energy terms are not so essential in the simulations of short (up to $n = 6$) alkanes where the molecular chains are almost rigid. On the contrary, for intermediate length alkanes (as well as for all long, flexible molecules like polymers or proteins) the 1-4 interaction scaling parameter has to be carefully adjusted, according to available experimental information. So, in principle, the scaling factor is a material-dependent characteristic in the current force fields.

The above problem is part of a more general question: how to account for electrostatic interactions. In the UA model the charges are set to zero. In the all-atom model there is a dipole on each C-H bond. However, since CHARMM does not take into account any polarizability of the molecules, the electrostatic energy must (somehow) account for these effects in an effective way. As we show below, the melting transition in tetracosane layers is extremely sensitive to both the 1-4 scaling factor as well as the distribution of effective charges.

Figure 2 presents the temperature dependence of the intermolecular van der Waals component of energy for three situations described above. For all cases, an abrupt change of the slope marking the melting transition is observed. However, the melting temperature is strongly dependent on the interaction model and varies in a very wide range, from 220 K to 450 K. According to experimental neutron diffraction data[9] a tetracosane monolayer adsorbed on graphite melts at approximately $T_M = 340$ K. The same value has been obtained in MD simulations using UA model of molecules[13,14]. However, when the all-atom model of molecule is used, the melting temperature is much lower, $T_M = 220$ K. Obviously, the UA molecular model and all-atom approach are not equivalent.

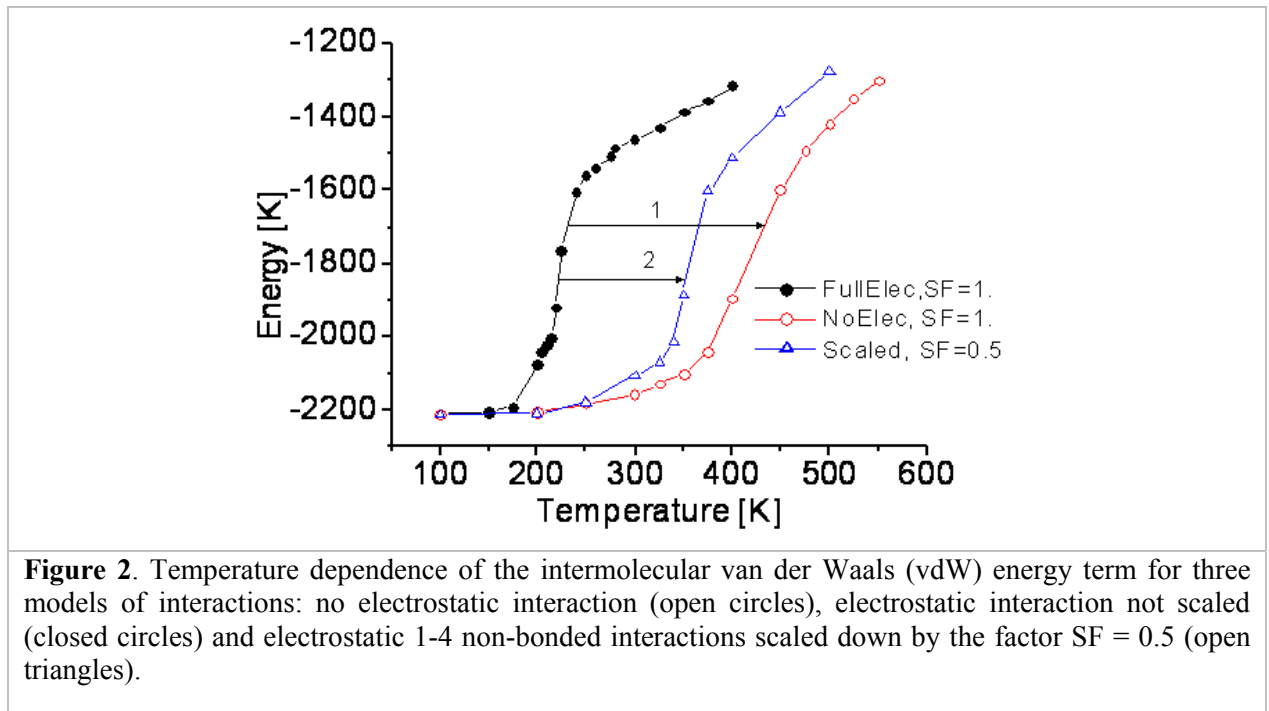

**Figure 2**. Temperature dependence of the intermolecular van der Waals (vdW) energy term for three models of interactions: no electrostatic interaction (open circles), electrostatic interaction not scaled (closed circles) and electrostatic 1-4 non-bonded interactions scaled down by the factor SF = 0.5 (open triangles).

Therefore, we assumed that if the electrostatic component of interactions were suppressed, $T_M$ should approach the experimental value. Indeed, the temperature of melting rose but to significantly higher value ($T_M = 450$ K). It meant that the direct application of the all-atom models of interaction (in two limits, with and without electrostatic energy) could not reproduce the experimental data and more subtle modification of interaction model was required. It could be done using the scaled 1-4 exclusion policy for non-bonded intramolecular interactions (as it should be for modern CHARMM force fields). So, we scaled the 1-4 non-bonded interaction and we found that for a scaling factor SF = 0.5 the melting temperature approached $T_M = 350$ K - a value that was close to the experimental one. The middle curve in the Fig. 2 represents this situation. We believe that such a scaling factor may compensate for the lack of polarizability inherent in the CHARMM force fields.

Figure 3 shows the temperature variation of the average fraction of *gauche* defects in the adsorbate molecules for all three models of interactions. All curves show an inflexion point at the same temperatures that the corresponding graphs of vdW energy (Fig.2), implying that the mechanism of melting is driven mainly by the internal bending of molecules. At the same time, the overall stiffness depends on the way of scaling of the 1-4 electrostatic energy. Large electrostatic contributions make the molecules softer. As a result, the *gauche* defects are formed at lower temperature and melting temperature decreases.

With increasing temperature, the deformation of molecules progresses from their ends towards their central part. Most internal bonds start to form *gauche* defects only at melting. Figure 4 shows the temperature variation of the fraction of *gauche* defects calculated as a function of dihedral angle index (21 angles in the chain of 24 carbon atoms) for the scaled interaction model.

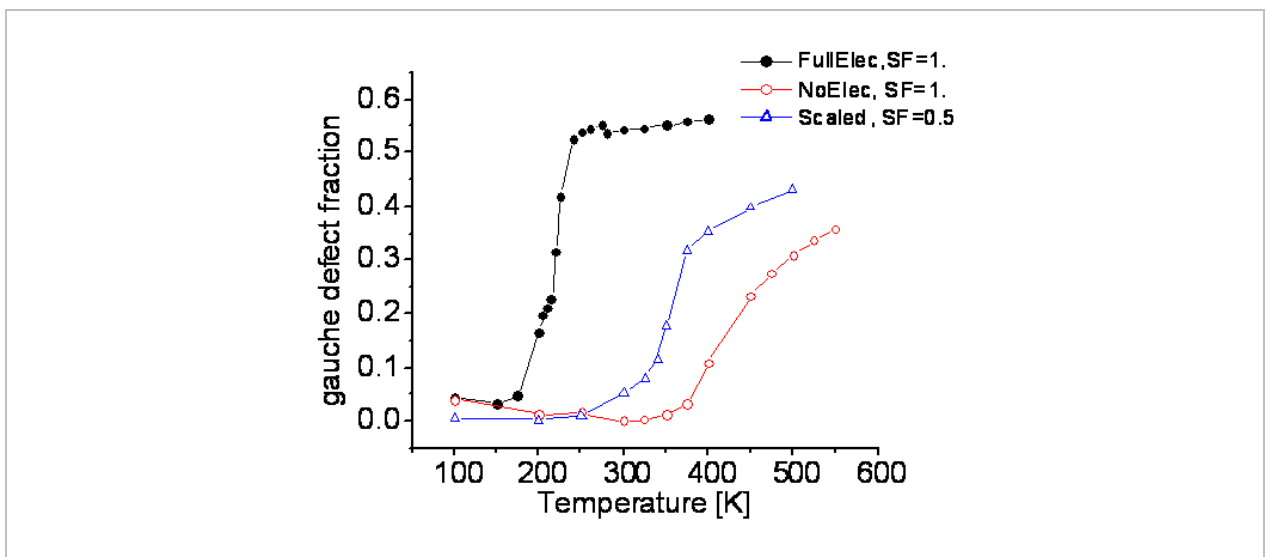

**Figure 3.** Fraction of bonds in gauche configuration as a function of temperature for three interaction models. The format is identical to that for Figure 2.

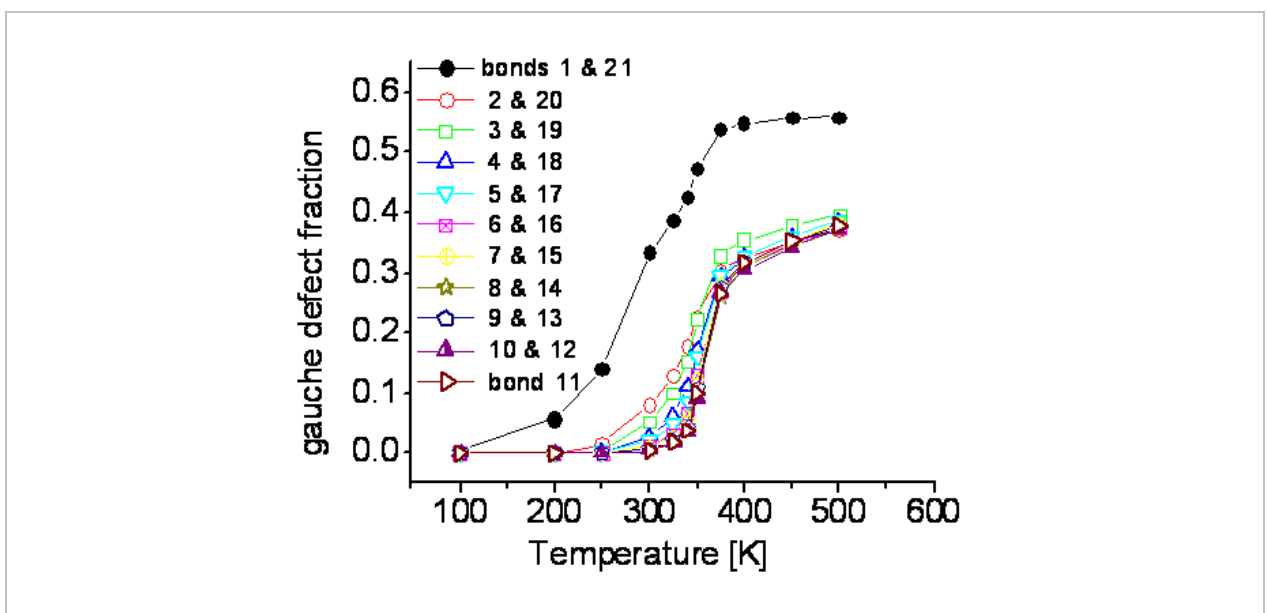

**Figure 4.** Fractions of gauche defects for different C-C bonds as a function of temperature for the scaled (SF = 0.5) interaction model. The averages were taken for the bonds located at the same distance from the ends of molecule. Bond 11 is located in the middle of molecule

    Similar behavior was observed for two other models of interaction. One can easily see, that the ends of molecules start to deform already at temperatures far below the melting ($T \sim 200K$). This behavior is the origin of the solid-smectic transition within the adsorbed film. We discuss mechanism of this transition in the next section of this paper.

    To evaluate the average changes in the molecular shape upon *gauche* defect formation, we have monitored the average end-to-end distance within molecules (Figure 5) as well as the end-to-end length distribution (Figure 6). The end-to-end distance is defined as the longest dis-

tance between two atoms in the same molecules, regardless of the type of atoms (carbon or hydrogen). As would be expected, the average molecular size decreases at melting (Fig.5). Such a size decrease is a consequence of the increasing number of *gauche* defects and the progressive change of linear shape of molecules towards a more globular form (but still confined to the plane). The strong holding potential of graphite prevents desorption of large segments of the molecule during bending and any formation of 3D globular-like species. On the other hand, it is important to remember that the shape of molecules in the melted (liquid) phase is not static; the instantaneous end-to-end distance is strongly fluctuating with an average amplitude ± 10 Å, as shown in Figure 6.

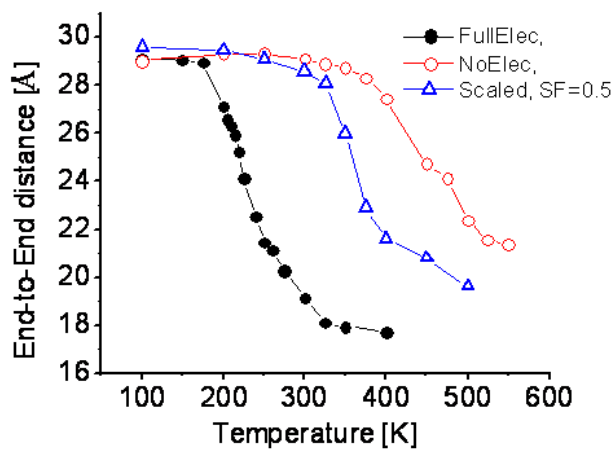

**Figure 5.** Average end-to-end molecular size as a function of temperature, for three interaction models.

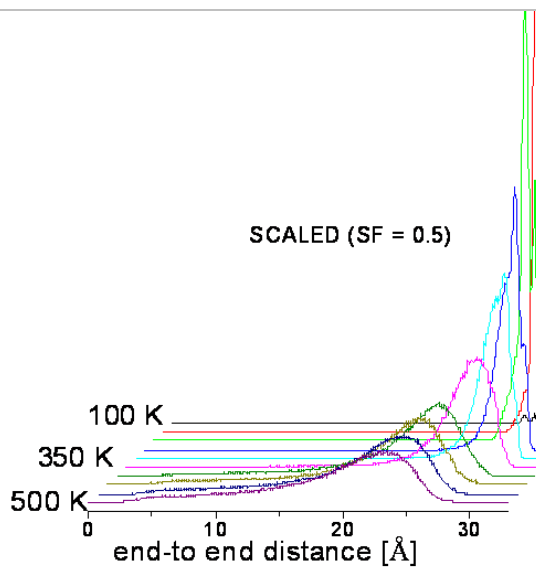

**Figure 6.** Distribution of the molecular end-to-end distance as a function of temperature for the scaled (SF = 0.5) interaction model.

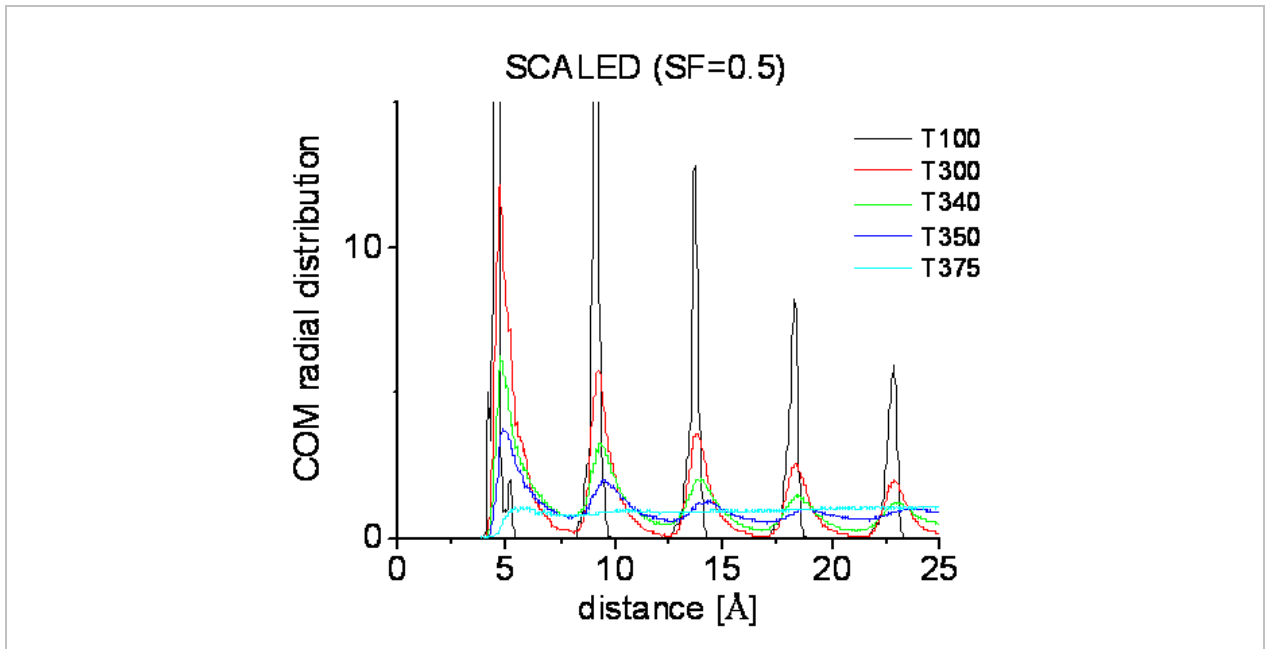

**Figure 7.** Pair correlation function of center of molecular masses for temperatures 100 K, 300 K, 340 K, 350 K and 375 K, for scaled (SF = 0.5) interaction model.

To conclude the discussion of melting, Figure 7 shows the pair correlation function *g(r)* calculated for the location of the molecular centers of mass for the scaled model of interaction and at different temperatures. Evidently, at temperatures where the tetracosane molecules lose their linear symmetry the traditional translational disordering of molecules also occurs. At low temperature ($T = 100$ K) the structure is solid. It becomes clearly liquid above $T = 350$ K. In the intermediate range of temperatures, a partial order persists, indicating that the phase behavior of the layer is more intricate. We address the mesophase in the following subsection. We limit the discussion to the interaction model that uses scaled 1-4 nonbonding energy term, as it reproduces the most closely the experimental features of melting, in particular, the melting temperature.

**2. Solid-smectic-liquid transformations.**

Structural properties of molecular layers are conveniently characterized by parameters which are sensitive to deformations of molecular arrangements with respect to an ideal symmetry. The order parameters $\Phi_n$ are defined as:

$$\Phi_n = \left| \frac{1}{N_b} \sum_{k=1}^{N_b} \exp(in\theta_k) \right|. \qquad (1)$$

$\Phi_n$ measures the average bond order within a plane layer. Each nearest neighbor position is characterized by a vector (bond) which has a particular orientation in the plane and can be described by the polar coordinate $\theta_k$. The parameter $n$ accounts for layer symmetry and gives the number of nearest neighbors in the ideal structure (for triangular lattice $n = 6$; for a square lattice $n = 4$). The index k runs over the total number of nearest neighbor bonds $N_b$ in the adsorbed layer. For a 2D solid with ideal translational order, $\Phi_n = 1$; whereas $\Phi_n = 0$ when the state of adsorbed layer is a two-dimensional isotropic fluid.

In order to quantify the orientational order in the system (and thereby the nematic character of the structure) we also monitored the nematic order parameter $OP_{nem}$. It was defined by Peters *et al.*[15,16] in the following way:

$$OP_{nem} = \left| \frac{1}{N_m} \sum_{i=1}^{N_m} \cos 2(\phi_i - \phi_{dir}) \right| \qquad (2)$$

Here $\phi_i$ is the angle that the axis of molecule (*i*) makes with the *arbitrary* chosen axis and $\phi_{dir}$ is a director of the average orientation of the system with respect to this axis.[17,18] Here $N_m$ is the number of molecules in the simulation box. For a perfect nematic, when all molecules have the same orientation, $OP_{nem} = 1$. In an isotropic fluid, where the orientational correlation between molecules is lost, $OP_{nem} = 0$.

Using $\Phi_n$ (*n* = 4 and 6) and $OP_{nem}$ parameters we detected some unusual features of the tetracosane monolayer dynamics at temperatures below melting. Figure 8 shows the temperature dependence of two of the $\Phi_n$ parameters, $\Phi_4$ and $\Phi_6$. According to this picture, the solid-liquid transformation goes through several well-distinguished steps.

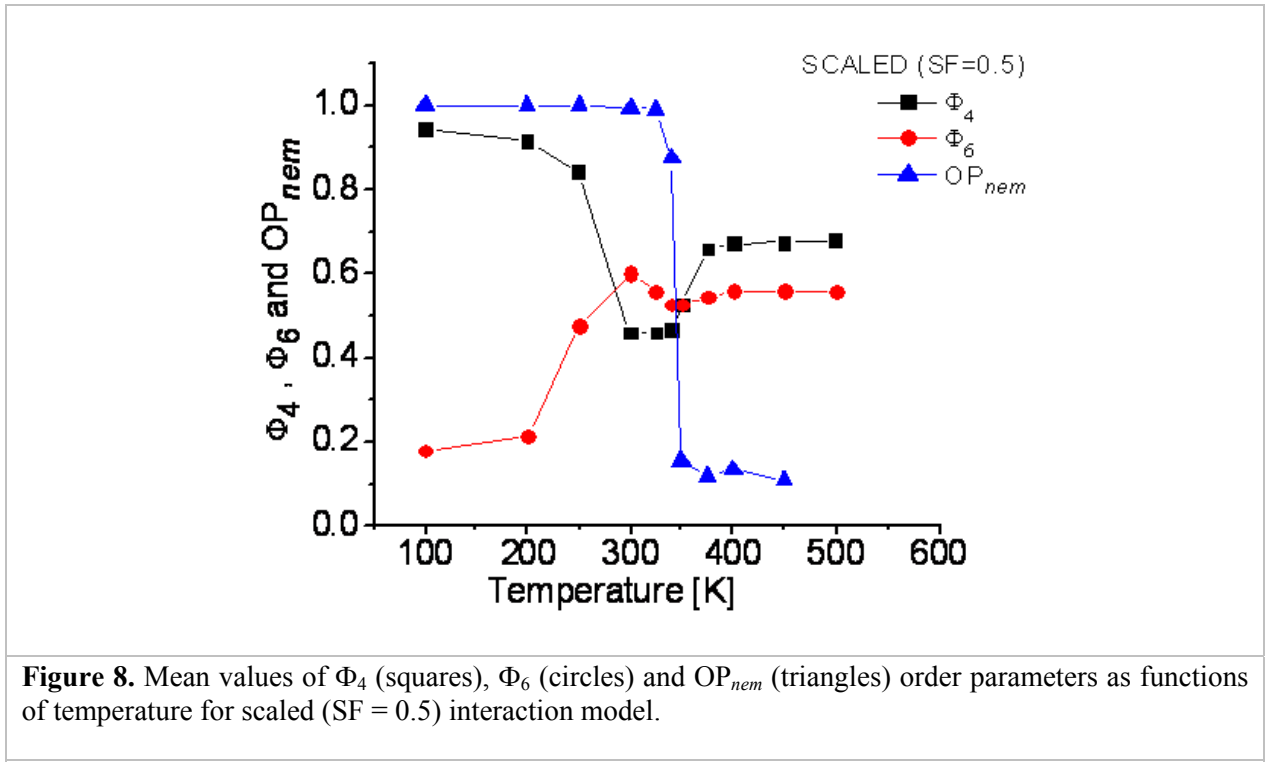

**Figure 8.** Mean values of $\Phi_4$ (squares), $\Phi_6$ (circles) and $OP_{nem}$ (triangles) order parameters as functions of temperature for scaled (SF = 0.5) interaction model.

Below *T* = 200 K the layer is obviously solid (see also Fig.9a and b). The value of $\Phi_4$ parameter is close to 1, indicating that the layer structure is rectangular. This is a consequence of the linear molecular structure and a 2D rectangular centered lattice with unit dimensions *a* = 4.53 Å, *b* = 64.6 Å defined by the positions of molecular centers of mass. For our computational system in the low-temperature solid phase the values of order parameters are: $\Phi_4 = 0.97$ and $\Phi_6 = 0.28$. The values observed in simulations agree fairly well with the theoretical ones (Fig. 8).

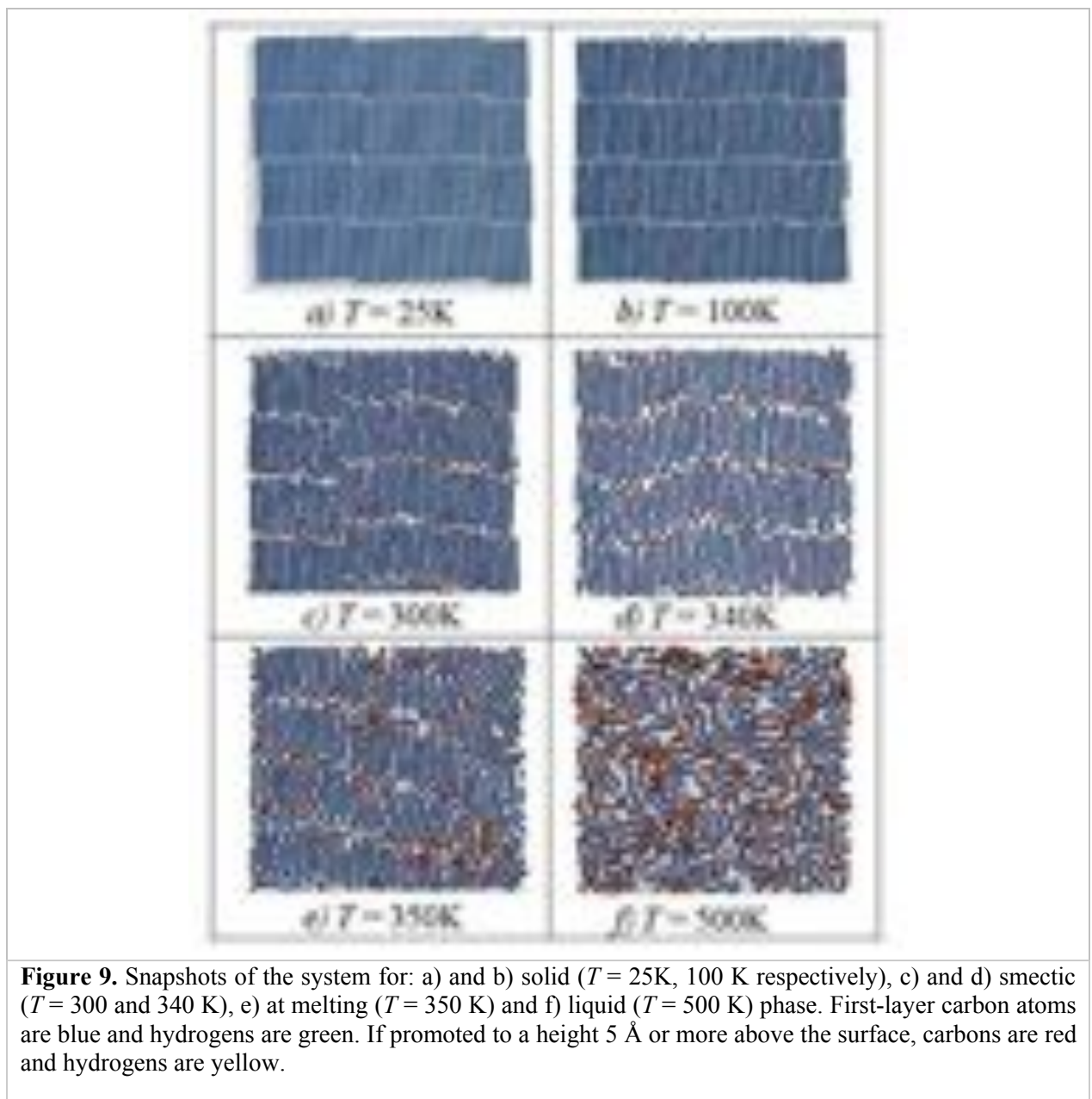

**Figure 9.** Snapshots of the system for: a) and b) solid ($T$ = 25K, 100 K respectively), c) and d) smectic ($T$ = 300 and 340 K), e) at melting ($T$ = 350 K) and f) liquid ($T$ = 500 K) phase. First-layer carbon atoms are blue and hydrogens are green. If promoted to a height 5 Å or more above the surface, carbons are red and hydrogens are yellow.

Above $T$ = 200 K the order parameters change rapidly when temperature goes up: $\Phi_4$ decreases and $\Phi_6$ increases, simultaneously. Such behavior signifies that the initial rectangular structure starts to deformed towards a more isotropic one. However, it is important to emphasize that the molecules themselves preserve their overall linear form, as the number of *gauche* defects is still very small (see Fig. 3) and the average end-to-end molecular size has only slightly changed (Fig. 5). The fluctuations of molecular positions lead to a displacement and broadening of peaks in the center-of mass pair correlation function *g(r)* (Fig. 7). Figure 10 shows a possible scenario of structural deformation which matches the order parameter behavior: the initial rectangular structure transforms towards "more triangular" arrangement and the $\Phi_6$ order parameter increases.

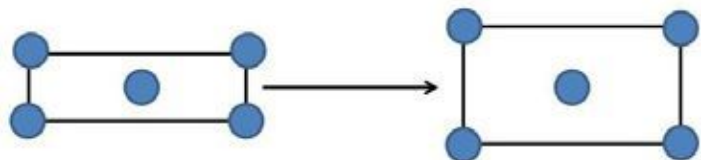

**Figure 10.** A scheme of structural deformation of rectangular centered 2D lattice of tetracosane at $T >$ 200 K.

So, it seems that in the 200–300 K temperature range the driving force of structural transformation is the translational instability of the system induced by a slight deformation of molecules due to *gauche* defect formation at their ends. As a consequence, the translational order is partially lost but the smectic-like arrangement of molecules in lamellae is not affected (Fig. 9c,d) and the nematic order parameter remains close to 1. The computational results are consistent with experimental neutron scattering data[13] that show that the low temperature rectangular structure undergoes a transition to a smectic-like structure at $T \sim 215$ K. According to the authors, in the smectic phase, although the molecules remain arranged in lamellae and are mostly parallel, some translational disorder is present. A similar picture also emerges from our all-atom simulations. To the best of our knowledge, the solid-smectic transition for the C24/graphite system was never before observed in a numerical model. In particular, it was not reproduced in recent UA simulations[13,14]. Such a result emphasizes the fact that in the case of modeling flexible molecules with explicit hydrogens it is extremely important to implement an accurate model of intramolecular interactions involving charge distribution and scaling of 1-4 nonbonding terms.

Above $T = 300$ K, $\Phi_4$ and $\Phi_6$ order parameters reverse their trends. The internal deformation of molecules starts to drive the transition, which is reflected by the increase of the number of *gauche* defects now occuring along the whole length of the molecule. As a consequence, the molecules lose orientational order and the lamellar structure of the layer progressively disappears (Fig. 9e). The average end-to-end size of molecules decreases and the center-of-mass distance from the substrate increases. The molecules become more globular in shape and much freer to rotate (Fig. 11). The nematic order parameter decreases to zero, indicating the absence of orientational order. At the same time, the $\Phi_n$ order parameters do not vanish but rather above T =340 K they both stabilize in the range 0.5–0.7. Such behavior indicates that some local translational order remains in the liquid phase (Fig. 9f) and that the melted tetracosane monolayer is not isotropic.

**Conclusions**

The simulations presented here give new insight into the structures and transformations in tetracosane ($C_{24}H_{50}$) monolayers adsorbed on graphite. The important conclusions summarized below are supported by good agreement with experimental observation[9,19].
  *(i)* The solid-smectic structural transition within the tetracosane monolayer has been observed in computer simulation for the first time. The proposed transition mechanism relates a loss of long-range translational symmetry to the local deformations of molecular linear chains. The *gauche* defects at the ends of molecules appear already at relatively low temperature (above ~ 200 K) and they induce the transition into the smectic structure. It is important to emphasize that this transition is very subtle and could be observed only in the variations of the $\Phi_n$ order parameters. The realistic representation of the tetracosane structure and a careful adjustment of the intermolecular interaction model also play an important role.

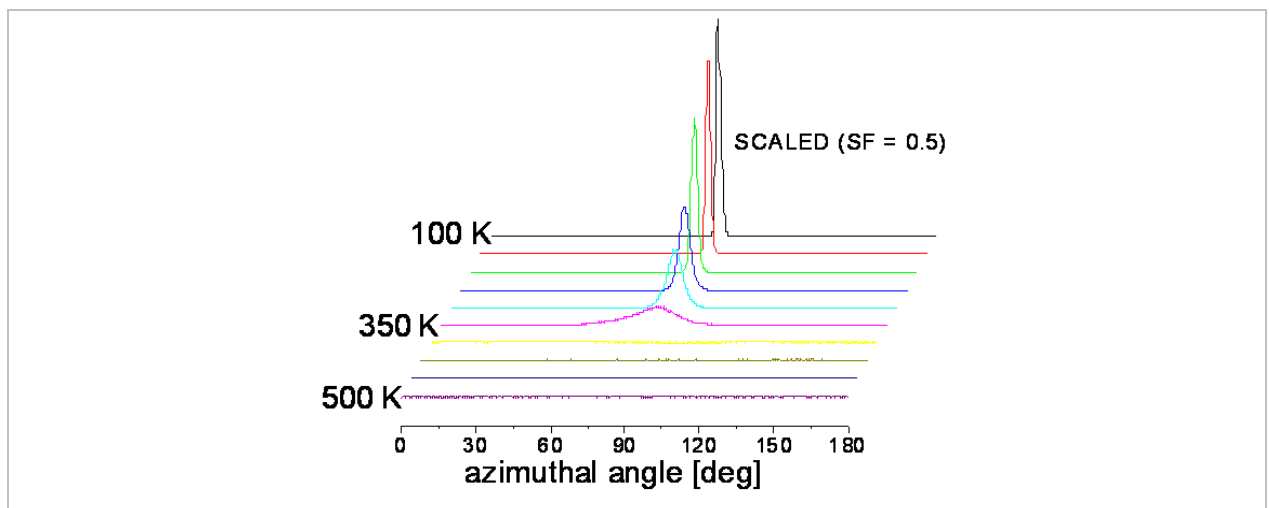

**Figure 11.** Azimuthal angle probability distribution for $T$ = 100, 200, 300, 350 and 375 K. Note that the unique peak at low temperatures characterizes the unique orientation of molecules with respect to the arbitrary chosen axis. At high temperature, in melted phase, the distribution is perfectly uniform.

*(ii)* The melting transition is strongly correlated with both internal (formation of the gauche defects) and rotational dynamics. The role of the molecules' deformation via modification of dihedral angles has already been stressed before[14]. Our simulations show that at melting the molecules assume a more globular shape which allows them to have more rotational freedom. However, the analysis of the $\Phi_n$ order parameters suggests that the final fluid preserves some degree of short-range translational order in agreement with the neutron diffraction pattern[9]. The intricate structure shown in the simulation snapshots in Fig. 11f is a consequence of the correlation between intra-chain and lattice melting in a constrained 2D geometry.

*(iii)* Both explicit representation of hydrogen atoms and a scaled all-atom potential were needed to reproduce the experimentally observed features. We showed that the temperature of the melting is extremely sensitive to the electrostatic interactions. We determined an appropriate scaling factor for the 1-4 non-bonded interactions.

**Acknowledgements:**

Acknowledgment is made to the Donors of The American Chemical Society Petroleum Research Fund (PRF43277 – B5), and the University of Missouri Research Board, for the support of this research. This material is based upon work supported in part by the Department of Energy under Award Number DE-FG02-07ER46411. The authors acknowledge useful discussions with Haskell Taub.